\documentclass[twocolumn,pra,showpacs,preprintnumbers,superscriptaddress]{revtex4}

\usepackage{graphicx}
\usepackage{dcolumn}
\usepackage{bm}

\begin{document}

\title{A Unified Theory of Consequences of Spontaneous Emission in a $\Lambda$ System}

\author{Sophia E. Economou}
\author{Ren-Bao Liu}
\author{L.J. Sham}
\affiliation{Department of Physics, University of California San
Diego, La Jolla, California 92093-0319}
\author{D.G. Steel}
\affiliation{The H. M. Randall Laboratory of Physics,
            University of Michigan,
            Ann Arbor, MI 48109}
\date{\today}

\begin{abstract}

In a $\Lambda$ system with two nearly degenerate ground states and
one excited state in an atom or quantum dot, spontaneous radiative
decay can lead to a range of phenomena, including electron-photon
entanglement, spontaneously generated coherence, and two-pathway
decay. We show that a treatment of the radiative decay as a
quantum evolution of a single physical system composed of a
three-level electron subsystem and photons leads to a range of
consequences depending on the electron-photon interaction and the
measurement. Different treatments of the emitted photon channel
the electron-photon system into a variety of final states. The
theory is not restricted to the three-level system.

\end{abstract}



\maketitle \normalsize

\section{Introduction}

The electromagnetic vacuum is commonly considered as a reservoir
which causes decoherence and decay of a quantum mechanical system
coupled to it. An alternative view holds that the two subparts
(`quantum system' and `bath') are constituents of a single closed
quantum mechanical whole, which is governed by unitary evolution
until a projection (measurement) is performed. Different
projections may give rise to a variety of phenomena which on the
surface appear unrelated. Spontaneous emission is a quantum
phenomenon which has been treated in both ways. Its effects are of
interest from the views of both fundamental physics and
applications.

The radiative decay of a three-level system is attractive for its
simplicity and yet richness in physical phenomena.  A variety of
effects follow from the spontaneous decay.  Those which involve
semiclassical light and ensemble of atoms include the
electromagnetically induced transparency \cite{EIT} and lasing
without inversion \cite{LaserNoInv}. By definition, a $\Lambda$
system has two nearly degenerate ground states which are
dipole-coupled to one excited state for optical transitions. We
shall, for conciseness, refer to the states as electronic states
in an atom or quantum dot. The decoherence and decay effects for a
single $\Lambda$ system  are relevant to  quantum computating and
information processing, for example in many implementation schemes
\cite{Imamoglu_CQED_Spin, OpticalControl_Sham, monroe_atom,
OpticallatticeQC,ReadWrite}, which can be more practical than the
direct excitation of the two-level system.

 A  $\Lambda$ system initially in the excited state will eventually decay by the emission of a photon.
This process may result in the entanglement of the $\Lambda$
system with the emitted photon.  Recently, entanglement between
the hyperfine levels of a trapped ion and the polarization of a
photon spontaneously emitted from the ion was demonstrated
experimentally \cite{monroe}.

In quantum optics of the atom, coupling to the modes of the
electromagnetic vacuum can contribute to coherence between atomic
states, and such terms have been implicit in the textbook
treatment of spontaneous radiative decay \cite{tannoudji} or
indeed explicit in research papers \cite{cardimona}.  In the early
90's, it was pointed out that in a $\Lambda$ system the
spontaneous decay of the highest state to the two lower ones may
result in a coherent superposition of the two lower states
\cite{Java}.  The conditions for this Spontaneously Generated
Coherence (SGC) as presented in Ref.~\onlinecite{Java} are that
the dipole matrix elements of the two transitions are
non-orthogonal and that the difference between the two frequencies
is small compared to the radiative line-width of the excited
state.

The final example is the so-called two-pathway decay in which a $\Lambda$-system --as
opposed to a $V$ system-- cannot exhibit quantum beats
 because  the information on which decay
path of the system is in principle available by detection of the
atom, and therefore no beats are expected (p.~19 of Ref.~\onlinecite{quantumoptics}).

All the phenomena listed above, when viewed separately,  appear
unrelated, if not downright contradictory. In fact, they stem from
the same process, namely the radiative spontaneous decay of a
$\Lambda$-system. The primary purpose of this paper is to show how
they naturally emerge from the same time evolved composite state
of the whole system ($\Lambda$ subsystem and the electromagnetic
modes). From this treatment follow the conditions for each effect
in terms of the electron-photon coupling and in terms of different
ways of projecting the photon state by measurement. We also show
how a change of symmetry of the system by the introduction of a
perturbation may determine whether a  SGC will occur or not.

The second goal of this work is to analyze these effects in the
solid state, where the two lower levels of the $\Lambda$ system
are the spin states of an electron confined in a semiconductor
quantum dot. For this system, SGC has been given a theoretical
analysis and experimental demonstration \cite{Gurudev}, and we
further propose here an experiment for the demonstration of
spin-photon polarization entanglement. In our treatment, we
distinguish between a single system and an ensemble for the
various phenomena; in this context, we make a comparative study of
the solid state and the atomic system.

This paper is organized as follows:  In section~\ref{unified} we
present the time evolution of the decay process which leads to the
conditions for the occurrence of each of the listed phenomena. In
section~\ref{symsgc} we deduce a set of conditions on the symmetry
of the system for SGC. Sections~\ref{examplesat} and
\ref{examplesss} illustrate these conditions by specific examples
from atomic and solid state systems, respectively. We also present
the theory of the pump-probe experiment and derive the probe
signal, which is altered by the SGC term (section \ref{ppexp}).

\section{Spontaneous Emission as Quantum Evolution} \label{unified}

Consider a single $\Lambda$ system in a photon bath with modes
$|k\rangle$, where $k=(\mathbf{k},\sigma)$, $\mathbf{k}$ being the
wave vector and $\sigma$ the state with the polarization vector
$\boldsymbol{\varepsilon}_\sigma$. In the dipole and rotating-wave
approximation, the Hamiltonian for the whole system is given by
\begin{eqnarray}
H&=&\sum_k \omega_kb^{\dag}_kb_k +\sum _{i=1}^3\epsilon_i
|i\rangle\langle i| +\sum_{k; i=1,2 } g_{ik} b^{\dag}_k|i\rangle\langle
3| \nonumber \\ && \ \ \ \  +\sum_{k; i=1,2}g^*_{ik} b_k
|3\rangle\langle i|  ,
\end{eqnarray}
where $b_k$ destroys a photon of energy or frequency $\omega_k$
($\hbar =1$) and $|i\rangle$ is the electronic state with energy
or frequency $\epsilon_i$. The coupling between the photon and the
electron is $g_{ik}\propto  \boldsymbol{\varepsilon}_\sigma \cdot
\mathbf{d}_i$, where $\mathbf{d}_i$ is the dipole matrix element
for the transition $3\leftrightarrow i$. The $\Lambda$ system is
taken to be at $t=0$ in the excited level $|3\rangle$ (which can
be prepared by a short pulse), and the photon bath is in the
vacuum state, i.e., the whole system is in a product state. For
$t>0$, the composite wavepacket can be written as
\begin{equation}
|\psi(t)\rangle\equiv c_3(t)|3\rangle |\text{vac}\rangle +\sum_k
c_{1k}(t)|1\rangle|k\rangle +\sum_k c_{2k}(t)|2\rangle|k\rangle,
\end{equation}
where $|\text{vac}\rangle$ is the photon vacuum state. Evolution
of this state is governed by the Schr\"{o}dinger equation.

By the Weisskopf-Wigner theory  \cite{weiss_wigner} of spontaneous
emission \cite{quantumoptics}, the coefficient $c_3$ is obtained
by one iteration of the other coefficients:
\begin{eqnarray}
\partial_t c_3&=&-i\epsilon_3 c_3  -\sum_k
\left|g_{1k}\right|^2\int^{t}_0
e^{-i(\epsilon_1+\omega_k)(t-t')}c_3(t') dt' \nonumber
\\ && -\sum_k \left|g_{2k}\right|^2\int^{t}_0
e^{-i(\epsilon_2+\omega_k)(t-t')}c_3(t') dt'.
\end{eqnarray}
Since the electron--photon coupling is much weaker than the
transition energy in the $\Lambda$ system, the integrals in the
equation above can be evaluated in the Markovian approximation,
resulting in:
\begin{eqnarray}
 \partial_t c_3\approx -i\epsilon_3c_3-{\Gamma_{31}\over
2}c_3 -{\Gamma_{32}\over 2}c_3,
\end{eqnarray}
where
\begin{equation}
\Gamma_{3i}=2\sum_k \left|g_{2k}\right|^2\int^{t}_0
e^{-i(\epsilon_i+\omega_k)(t-t')} dt'.
\end{equation}

Thus, the solution is
\begin{eqnarray}
c_3\approx e^{-(i\epsilon_3 +\Gamma/2)t},
\end{eqnarray}
where $\Gamma\equiv\Gamma_{31}+\Gamma_{32}$ is the radiative
linewidth of the excited state. Furthermore, $c_{1k}$ and $c_{2k}$
are given by
\begin{eqnarray}
 c_{1k}\approx-{g_{1k}\over \epsilon_3-\epsilon_1-\omega_k-i{\Gamma\over 2}}
   \left[e^{-i(\epsilon_1+\omega_k)t}-e^{-i\epsilon_3 t-{\Gamma\over 2} t}\right], \nonumber \\
c_{2k}\approx-{g_{2k}\over
\epsilon_3-\epsilon_2-\omega_k-i{\Gamma\over 2}}
   \left[e^{-i(\epsilon_2+\omega_k)t}-e^{-i\epsilon_3
   t-{\Gamma\over 2}
   t}\right]. \nonumber
\end{eqnarray}
In order to study the system in the $2\times2$ subspace of the
lower states, we take the limit $t\gg \Gamma^{-1}$. After the
spontaneous emission process, the final state is a
electron--photon wavepacket $\sum_{k;
i=1,2}c_{ik}|i\rangle|k\rangle ,$ with the coefficients
\begin{eqnarray}
 c_{1k}\approx-{g_{1k}\over
 \epsilon_3-\epsilon_1-\omega_k-i{\Gamma\over 2}}
   e^{-i(\epsilon_1+\omega_k)t}, \label{c1}\\
c_{2k}\approx-{g_{2k}\over
\epsilon_3-\epsilon_2-\omega_k-i{\Gamma\over 2}}
  e^{-i(\epsilon_2+\omega_k)t}. \label{c2}
\end{eqnarray}
The state of a photon is specified by its propagation direction
$\mathbf{n}$, polarization $\sigma$
($\boldsymbol{\varepsilon}_\sigma\perp\mathbf{n}$), and frequency
$\omega$. So we can formulate the total wavepacket as
\begin{equation}
\sum_{\mathbf{n},\sigma} \left[g_{1\sigma}e^{-i\epsilon_1
t}|1\rangle|\mathbf {n},\sigma, f_1(t)\rangle
+g_{2\sigma}e^{-i\epsilon_2 t}|2\rangle|\mathbf {n},\sigma,
f_2(t)\rangle\right],\nonumber\\ \label{packet}
\end{equation}
where we have taken the coupling constants to be
frequency-independent. In Eq.~(\ref{packet}) $f_j(t)$ is the pulse
shape of the photon. From Eq.~(\ref{c1}) and (\ref{c2}), we see
that the photon wavepacket has a finite bandwidth; this point,
which was first studied by Weisskopf and Wigner in their classic
treatment of spontaneous emission \cite{weiss_wigner}, is
reflected in the structure of $f_{j}(t)$. These functions have a
central frequency equal to $\epsilon_3 -\epsilon_{j}$ and a
bandwidth equal to $\Gamma$. As a consequence of the finite
bandwidth, for a given propagation direction and polarization, the
basis states $\{|\mathbf{n},\sigma,f_j\rangle\}$ are not
orthogonal, the overlap between them being
\begin{equation}
\langle \mathbf{n},\sigma, f_l|\mathbf{n},\sigma,f_j\rangle =
\frac{i\Gamma}{i\Gamma+\epsilon_{lj}},
\end{equation}
where $\epsilon_{lj}=\epsilon_l -\epsilon_{j}$.

We should emphasize that the wavepacket formed in
Eq.~(\ref{packet}) does not rely on the Markovian approximation.
In a full quantum kinetic description of the photon emission
process, the wavepacket of the whole system would still have the
same form, the central frequency and bandwidth of the pulses would
be close to those found using the Markovian approximation, but the
specific profile of $f_j(t)$ would be different from those given
by Eq.~(\ref{c1}) and (\ref{c2}).

The various phenomena (electron and photon polarization
entanglement, SGC, and two-pathway decay) can all be derived from
the wavepacket of Eq.~(\ref{packet}).

If the spontaneously emitted photon is not detected at all, we
have to average over the ensemble of photons of all possible
propagation directions to obtain the electronic state. This is the
usual textbook treatment of spontaneous emission. However, if
detection  of an emitted photon leads to a knowledge that its
direction of propagation is $\mathbf {n}_0$, then the
(unnormalized) electron-photon wavepacket should be projected
along that direction:
\begin{equation}
\sum_{\sigma} \left[g_{1\sigma}e^{-i\epsilon_1 t}|1\rangle|\mathbf
{n}_0,\sigma,f_1(t)\rangle +g_{2\sigma}e^{-i\epsilon_2
t}|2\rangle|\mathbf
{n}_0,\sigma,f_2(t)\rangle\right]\!.\nonumber\\ \label{composite}
\end{equation}
When the two transitions are very close in frequency, i.e.,
$\eta\equiv|\epsilon_1-\epsilon_2|/\Gamma\ll 1$, the overlap of
the two photon wavepackets deviates from unity by $
\mathcal{O}(\eta)$. After tracing out the envelopes of the photon
by use of any complete basis (e.g. monochromatic states), the
state of the electron and photon polarization is, with the
propagation direction $\mathbf{n}_0$ understood,
\begin{eqnarray}
|\Upsilon\rangle = \sqrt{N} \sum_{\sigma} \left[g_{1\sigma}|1\rangle|\sigma\rangle
+g_{2\sigma}|2\rangle|\sigma\rangle\right] + \mathcal{O}(\eta) , \label{explicit}
\end{eqnarray}
where $N$ is a normalization constant, given by
\begin{equation}
N^{-1} =\sum_{j=1,2}\sum_{\sigma=\alpha,\beta}|g_{j\sigma}|^2.
\end{equation}
The order $\eta$ error recorded here is meant to indicate the
magnitude of the \textit{mixed-state} error which, if neglected,
results in a pure state. From this pure state, we can find
explicitly the necessary conditions for entanglement or SGC.
However, the approximation of neglecting $\eta$ is unnecessary for
computing a measure of entanglement of the resultant mixed state
\cite{bdsw}.

\subsection{Entanglement}

A measure of entanglement of the bipartite state
$|\Upsilon\rangle$ in Eq.~(\ref{explicit}) is given by the von
Neumann entropy of the reduced density matrix of the state
\cite{wootters} for either the subsystem $E$ of the two low-lying
electronic states or the subsystem $P$ of the photon polarization
states. Taking the partial trace of the polarization states of the
density matrix $|\Upsilon\rangle\langle\Upsilon|$ of the pure
state  leads to the $2\times2$ reduced density matrix for the
electronic states,
\begin{equation}
\rho_E = N \sum_{ij} |i\rangle \left[\sum_\sigma g_{i\sigma}
g^*_{j\sigma} \right] \langle j|.
\end{equation}
 Diagonalization of this partial density matrix leads to two eigenvalues,
 \begin{equation}
p_\pm = \frac{1}{2} \pm \sqrt{\frac{1}{4} - D^2},
\end{equation}
where $D^2$ is the determinant of the reduce density matrix $\rho_E$, or
 \begin{equation} \label{maxent}
D= N |g_{1\alpha}g_{2\beta}-g_{1\beta}g_{2\alpha}|, \label{discrim}
\end{equation}
for the two electronic state and two polarizations, $\alpha,\beta$, normal to the
propagation direction $\mathbf{n}_0$. The entropy of entanglement is given by the entropy,
\begin{equation}
S = - p_+ \log_2 p_+ - p_- \log_2 p_-
\end{equation}
As $D$ ranges from 0 to  $1/2$, the entropy ranges from 0 to 1
giving a continuous measure of entanglement as the state
$|\Upsilon\rangle$ goes from no entanglement to maximum
entanglement. To find the axis $\mathbf{n}_0$ along which the
entanglement is maximum, we have to maximize $D$ as a function of
the orientation. For a particular system, this axis can be found
in terms of the dipole matrix elements of the two transitions.
However, not all systems can have maximally entangled states. We
will apply this to specific examples in the following section.

\subsection{SGC}

From the reduced density matrix, we can also find the conditions
for SGC. Maximum SGC occurs when the reduced density matrix is a
pure state. In terms of the electron-photon coupling constants the
condition is the vanishing of the discriminant $D$ in
Eq.~(\ref{discrim}). This means that when the SGC effect is
maximized, there exists a particular transformation which takes
the basis of the electronic states $\{|1\rangle, |2\rangle\}$ to a
basis $\{|\mathcal{B}\rangle, |\mathcal{D}\rangle\}$ which has the
property that $|\mathcal{B}\rangle$ is always the final state of
the $\Lambda$-system immediately after the spontaneous emission
process,  and $|\mathcal{D}\rangle$ is a state disconnected from
the excited state by dipolar coupling, i.e. a dark state. This
point will be further explored in section \ref{symsgc}. The
extreme values of $D=0$ and $1/2$ make it clear that maximum SGC
means no entanglement and conversely that maximal entanglement
leads to no SGC. However, partial entanglement can coexist with
the potentiality of some SGC for values of $D$ between the two
extremes.

Our theory can be easily extended to systems with more than two
ground states. For example, in a system whose ground states are
the four states from two electron spins, the SGC may lead to the
coherence and entanglement between the two spins, which is the
mechanism of a series of proposals of using vacuum fluctuation to
establish entanglement between qubits
\cite{plenio_cavloss,entangle_dist_at}.

\subsection{Two-pathway decay}

So far we have investigated the consequences when the two
transitions are close in frequency ($\eta\ll 1$). When this is not
the case, the tracing-out of the wavepacket will generally produce
a mixed state in electron spins and photon polarizations. In the
limit of large $\eta$, i.e.,
$\left|\epsilon_2-\epsilon_1\right|\gg\Gamma$, the overlap between
the two photon wave functions, $\langle f_1(t)|f_2(t)\rangle\simeq
0$, and the reduced density matrix for the spin and photon
polarization would be mixed. In this case there is neither
spin-polarization entanglement nor SGC, but instead the time
development can be described as a two-pathway decay process: the
excited state can relax to two different states by the emission of
photons with distinct frequencies. For $\eta$ between these two
limits, the state in Eq.~(\ref{composite}) may lead to an
entanglement between the pulse shapes of the photon and the two
lower electronic levels on measuring the photon polarization.
Furthermore, from the entangled state in Eq.~(\ref{composite}),
SGC or polarization entanglement may still be recovered (provided
of course that the necessary conditions on the $g$'s are
satisfied) if  the quantum information  carried by the frequency
of the photon is erased \cite{quantum_eraser}. This can be done by
chopping part of the photon pulse, and thus subjecting its
frequency to (more) uncertainty. In a time-selective measurement,
only photons emitted at a specific time period, say from $t_o$ to
$t_o+dt$, are selected. So the projection operator associated with
this measurement is $P_o=|\delta(t-t_o)\rangle
\langle\delta(t-t_o)|$, which represents a $\delta$ photon pulse
passing the detector at $t=t_o$. The projected state after this
measurement
\begin{eqnarray}
\sum_{\sigma}
\left[g_{1\sigma}f_1(t_o) |1\rangle
 +g_{2\sigma} f_2(t_o) |2\rangle
\right]|\mathbf {n}_0 \sigma \rangle \label{projection}
\end{eqnarray}
is a pure state of the electron and photon-polarization, so that
entanglement  or SGC is restored. By writing the projector in the
frequency domain
\begin{equation}
\tilde P_o=\int d\omega \int d\omega'\, e^{i(\omega-\omega')t_o}|\omega'\rangle\langle\omega| ,
\end{equation}
we see that it can be understood as a broadband detector with definite phase for
each frequency channel; thus it can erase the frequency (which-path) information
while retaining the phase correlation. We note that a usual broadband
detector without phase correlation is not sufficient to restore the pureness
of the state. It is also interesting that
SGC and entanglement can be controlled by choosing a different detection time
$t_o$, as seen from Eq.~(\ref{projection}).

\section{Symmetry Considerations for SGC} \label{symsgc}

 In this section we investigate the symmetry relations
between the different parts of the Hamiltonian necessary for SGC
terms to appear.  Our treatment is not restricted to $\Lambda$
systems, but can be extended to a system with more than two lower
levels.

 Consider a quantum mechanical system with one higher energy level
$|e\rangle$ and a set of lower-lying states, described by a
Hamiltonian $H^o$.  Taking into account only dipole-type
interactions, denote by $\mathcal{J}_z$ the polarization operator
used in the selection rules.  The $z$ axis is defined by the
excited state via
\begin{displaymath}
\mathcal{J}_z|e\rangle = M_e |e\rangle
\end{displaymath}
Note that $\mathcal{J}_z$ can be either $J_z$, where
$\bm{J}=\bm{L}+\bm{S}$ is the total angular momentum operator and $\bm{S}$ is the spin, or
$L_z$, as determined by the condition
\begin{displaymath}
[\mathcal{J}_z,H^o] = 0.
\end{displaymath}
That is to say there is an axial symmetry in the system associated
with $\mathcal{J}_z$.  Among the lower lying states, the ones of
interest are the ones appearing in the final entangled state
$|\Upsilon\rangle$ of the whole system. We will refer to these
states as `bright', because they are orthogonal to the familiar
dark states from quantum optics. There are at most three such
states, $\{\mathcal{B}_j\}$, within a given degenerate manifold,
corresponding to the three different possible projections of the
dipole matrix elements along the $z$ axis, so $j=1,0,\bar{1}$. In
general, not all systems will have all three bright states. This
concept that the final state involves only a small number of
states (three in our case), gives a physical understanding of the
electron-photon entangled state \cite{eberly}.

 In order to have SGC,
i.e., one or more terms of the type
$\dot{\rho}_{jk}=\Gamma\rho_{ee}$, with $j\neq k$  and $j,k\neq
e$, there has to be a perturbation $H^B$ that breaks the symmetry
associated with $\mathcal{J}_z$; in particular, the following
conditions have to be satisfied:
\begin{enumerate}
\item[(i)] $[H^B,\mathcal{J}_z]\neq 0$;
\item[(ii)] $H^B|e\rangle\propto|e\rangle$;
\item[(iii)] $|\epsilon_{12}|\lesssim\Gamma$.
\end{enumerate}
In general, we expect SGC between two eigenstates of the
Hamiltonian $H=H^o+H^B$ which have nonzero overlap with the same
bright state. The role of the first condition is to make SGC
non-trivial; without this condition, it would always be possible
to rotate to a different basis and formally acquire an SGC-like
term in the equations (e.g. by rotating to the $x$ basis in the
zero magnetic field case in the heavy-hole trion system discussed
below). The second condition ensures that the excited state will
not mix under the action of $H^B$; relaxing this condition gives
rise to the Hanle effect \cite{tannoudji,quantumoptics}, in which
an ensemble of atoms in a magnetic field is illuminated with an
$x$-polarized pulse and the reradiated light may be polarized
along $y$. This effect is another example where coherence plays an
important role; it has recently been observed in doped GaAs
quantum wells, in the heavy-hole trion system with confinement in
one dimension \cite{gammon}.
 We shall discuss the quantum dot case below.
As shown in Sec. \ref{unified}, when the radiative line-width of
the excited state is smaller than the energy differences of the
lower states the SGC effect will be averaged out. The third
condition provides the valid regime for the occurrence of this
phenomenon.

The perturbation $H^B$ can be realized by
a static electric or magnetic field, by the spin-orbit coupling,
by hyperfine coupling, etc.
 Note the different origins of $H^B$
in different systems and that it may or may not be possible to
control $H^B$.  Examples of various systems follow, exhibiting the
above conditions and demonstrating the different origins of $H^B$.

\section{Examples from Atomic Physics} \label{examplesat}

\subsection{SGC in atoms} Consider an atom
with Hamiltonian $H^o$; excluding relativistic
corrections, it can be diagonalized in the
$|N,L,S,M_L,M_S\rangle$ basis.  Consider as the system of interest
the subspace of $H^o$ formed by $|N,1,1,1,1\rangle=|e\rangle$ and
the lower-energy states $|N-1,L,S,M_L,M_S\rangle$.  The various
quantum numbers are of course restricted by selection rules, and
$\mathcal{J}_z=L_z$ . Here we will list only the three bright
states:
\begin{displaymath}
|\mathcal{B}_1\rangle=|N-1,2,1,2,1\rangle
\end{displaymath}
\begin{displaymath}
|\mathcal{B}_0\rangle=|N-1,2,1,1,1\rangle
\end{displaymath}
\begin{displaymath}
|\mathcal{B}_{\bar1}
\rangle=a|N-1,2,1,0,1\rangle+b|N-1,0,1,0,1\rangle
\end{displaymath}
where the coefficients $a$ and $b$ can be determined in the
following way: in the original $|NJM_JLS\rangle$ basis, the matrix
elements for the transitions
$|N-1,2,1,0,1\rangle\leftrightarrow|N,1,1,1,1\rangle$ and
$|N-1,0,1,0,1\rangle\leftrightarrow|N,1,1,1,1\rangle$ are given by
the Wigner-Eckart theorem. By rotating to the
$\{|\mathcal{B}\rangle,|\mathcal{D}\rangle\}$ basis, and requiring
the transition
$|\mathcal{D}\rangle\leftrightarrow|N,1,1,1,1\rangle$ to be
forbidden, we find $a$ and $b$.
  Inclusion of the spin-orbit interaction, which plays the
role of $H^B$, i.e. $H^B=\alpha \bm{L}\cdot\bm{S}$, condition (i)
is satisfied, the eigenstates of $H^B$ being
$|NJM_JLS\rangle$. Condition (ii) is also satisfied, because
$|e\rangle$, as the state of maximum $M_L$ and $M_S$, does not
mix under the spin-orbit coupling. In the new basis, SGC is
expected to occur between states with the same value of $M_J$,
which can also be verified by direct calculation. In this example
the line-width of $|e\rangle$ is much smaller than the spin-orbit
coupling strength $\alpha$. Typical
values in atoms are $\Gamma_e\sim 1\mu $~eV and $\alpha\sim 1 $~meV,
which means that SGC will not be observed in such a system.

\subsection{Entanglement and SGC of atomic hyperfine states}

In this example, the $\Lambda$ system is formed by the hyperfine
states of a single trapped Cd ion in the presence of a magnetic
field along the $z$ axis. In the $|FM_F\rangle$ basis, the excited
state is $|21\rangle$ and the two lower levels are $|11\rangle$
and $|10\rangle$. The two lower levels have the same principle
quantum number $N$. The entanglement between the polarization of
the photon and the atom has been demonstrated experimentally
\cite{monroe}. To illustrate the methods developed in Section
\ref{unified}, we will make use of the fact that the two lower
levels are states of definite angular momentum and its projection
to the $z$ axis. Then, by the Wigner-Eckart theorem we know that
the dipole moment of the transition
$|21\rangle\rightarrow|10\rangle$ has a nonzero component only
along $\mathbf{e}_+=\mathbf{x}+i\mathbf{y}$ whereas that of
$|21\rangle\rightarrow|11\rangle$ has only a component along
$\mathbf{z}$. The wavepacket of the system is then given by
\begin{equation}
|\Upsilon\rangle
=\frac{-\sqrt{2}\sin\vartheta|\vartheta\rangle|11\rangle
+e^{-i\varphi}\cos\vartheta|\vartheta\rangle|10\rangle -
ie^{-i\varphi}|\varphi\rangle|10\rangle}{\sqrt{2+\sin^2\vartheta}}
\label{psi}
\end{equation}
where $\vartheta$ and $\varphi$ are the spherical coordinates
measured from $z$ and $x$ axis, respectively, and
$|\vartheta\rangle$ and $|\varphi\rangle$ are the polarization
basis states, which are linearly polarized parallel and normal to
the plane formed by the $z$ axis and the propagation direction,
respectively. Then from Eq.~(\ref{psi}), we read off the $g$'s:
\begin{eqnarray}
g_{1\vartheta} &\propto& -\sqrt{2}\sin\vartheta  \\
g_{1\varphi} &=& 0 \\
g_{2\vartheta} &\propto& e^{-i\varphi}\cos\vartheta  \\
g_{2\varphi} &\propto& i e^{-i\varphi},
\end{eqnarray}
where $|11\rangle\equiv|1\rangle$ and $|10\rangle\equiv|2\rangle$.
The measure of entanglement by $D$ is
\begin{equation}
D = \frac{\sqrt{2}\sin\vartheta}{\sqrt{2+\sin^2\vartheta}}.
\label{trapi}
\end{equation}
The maximum possible entanglement occurs at $\vartheta=\pi/2$,
i.e., whenever the photon propagates perpendicularly to $z$. The
maximum value of 0.47 is close to being maximally entangled. $D$
does not depend on $\varphi$, as expected since there is azimuthal
symmetry about $z$.

 In terms of SGC and symmetry, it is interesting to notice that
the role of the (external or internal) field, $H^B$, introduced in
section~\ref{symsgc} can be played by the different projections
(measurements) because the state before the measurement is an
eigenstate of the operator $J_z$ (total angular momentum along
$z$) but not after the measurement in general.  The magnetic field
along the $z$-axis is included in the Hamiltonian $H^o$. If the
spontaneously emitted photons are measured along the quantization
axis, only the ones emitted from the transition
$|21\rangle\rightarrow|10\rangle$ will be detected, since only
their polarization allows propagation along $z$.  On the other
hand, a photon detector placed at a finite angle from $z$ can play
the role of $H^B$. Suppose a photon is spontaneously emitted along
an axis $n=(\vartheta,\varphi)$. The density matrix of the state
given by Eq.~(\ref{psi}) is $|\Upsilon\rangle\langle\Upsilon|$. If
we are only interested in the dynamics of the ion, and the
polarization of the photon is not measured, then the photon
polarization has to be traced out. Then the reduced density matrix
of the system, in the atomic states is
\begin{equation} \rho_E =\frac{1}{2+\sin^2\vartheta}\left[\begin{array}{cc}
\cos^2\vartheta+1 &
\sqrt{2}e^{-i\varphi}\cos{\vartheta}\sin{\vartheta} \\
\sqrt{2}e^{+i\varphi}\cos{\vartheta}\sin{\vartheta} &
2\sin^2{\vartheta}
\end{array} \right].
\end{equation}
The off-diagonal elements express coherence between the hyperfine
states with dependence on the photon propagation direction. We can
check that for $\vartheta=0$ the probability of the atom being in
the $|11\rangle$ state is zero and there are no off-diagonal
elements, and for $\vartheta = \pi/2$ the off-diagonal elements
are also zero, which means there is no SGC, but the state has the
maximum possible entanglement.  For all the intermediate values of
$\vartheta$ the hyperfine states and the photon polarization are
entangled, and there is also some SGC when the photon is traced
out. Maximum SGC occurs when $D$ is minimized; from
Eq.~(\ref{trapi}) we see that it is zero for $\vartheta=0$. This
is expected anytime the one of the two transitions involves a
linearly polarized photon, since the latter cannot propagate along
the quantization axis. So, for this orientation the final state
can only be $|10\rangle$. For intermediate angles, for instance
$\vartheta=\pi /4$, there is both entanglement and SGC involving
both lower states, when the photon is traced out. Since SGC only
occurs for particular photon propagation directions we could view
it as `probabilistic' SGC.

\section{Examples from Solid State Physics} \label{examplesss}

\subsection{Heavy-hole trion system in a magnetic field}

 In the optical control of the electron spin in a
doped quantum dot \cite{OpticalControl_Sham}, a static magnetic
field is imposed in a fixed direction at an angle $\psi$ with
respect to the propagation of the circularly polarized pulse along
the growth direction of the dot, defined as the $z$ axis. The two
eigenstates of the electron spin along the field direction and the
intermediate trion (bound state of an exciton with the excess
electron) state in the Raman process form a three-level $\Lambda$
system. The trion state of interest consists of a p-like heavy
hole and a pair of electrons in the singlet state.  The $g$-factor
in the $xy$ plane of the heavy hole is approximately zero in
magnetic fields up to 5 T \cite{Tischler_fineStruc} and the two
electrons are in a rotationally invariant state.  This means that
the trion state, although it is spin polarized along $z$, will not
precess about a perpendicular $B$-field. Therefore it can be
described by the `good' quantum numbers $J=3/2$ and its projection
along $z$, $M_J=3/2$. The lower levels $|1\rangle$, $|2\rangle$
are the eigenstates of the spin along the direction of the
$B$-field and have $j=1/2$ and $m_j=1/2,-1/2$ respectively.

To check if this system will have SGC, we will examine  whether
the conditions of Section \ref{symsgc} are satisfied. We take
$H^o$ to be the Hamiltonian of the Q.D., with
$|e\rangle=|\tau\rangle$, the trion state described above, excited
by $\sigma+$ light; $\mathcal{J}_z=J_z$, since the spin-orbit
interaction is included in $H^o$, and  any component of the $B$
field along $z$ can also be included. $H^B$ is the contribution to
the Hamiltonian due to the magnetic field along $x$. Condition (i)
is fulfilled since $g_x\simeq 0$, and condition (ii) is obviously
satisfied. The only bright state is the electron spin $s_z$
eigenstate, $|z\rangle\equiv|\uparrow\rangle$. For later use, we
also define $|\bar{z}\rangle\equiv|\downarrow\rangle$. Therefore
we expect SGC between states $|1\rangle$ and $|2\rangle$ for any
angle $\psi$, and since the linewidth of the trion is large enough
compared to the Zeeman splitting, SGC should moreover have a
detectable effect. As a matter of fact, it has already been
demonstrated experimentally for this system, and, to the best of
our knowledge, it is the only direct observation of SGC
\cite{Gurudev}. For this nonlinear pump-probe experiment, the
inclusion of SGC into the equations causes the amplitude and the
phase of the probe signal to depend on the Zeeman splitting. More
details on how this
dependence occurs will be presented in the following section.\\

Although our discussion has focused on single $\Lambda$ systems,
the experiment was carried out for an ensemble.  In general, for
an ensemble of equivalent non-interacting atoms, an average over
the different $z$ axes would have to be performed.  However, in
this quantum-dot solid state system, there is a common $z$ axis
for all the dots, since they are grown on the same plane ($xy$),
and they have a relatively large in-plane cross-section as
compared to their height. This is a clear advantage of the quantum
dot ensemble over an ensemble of atoms.

 We can also analyze this system using the methods in Section \ref{unified}.
To find the $g$'s, we need the dipole matrix elements. These can
be found by writing
\begin{eqnarray}
|1\rangle = \cos\frac\psi 2|\uparrow\rangle+ \sin\frac\psi 2|\downarrow\rangle\\
|2\rangle = \sin\frac\psi 2|\uparrow\rangle- \cos\frac\psi
2|\downarrow\rangle
\end{eqnarray}
Again, we will make use of the fact that $|\uparrow\rangle$ and
$|\tau\rangle$ are angular momentum eigenstates along the $z$
axis, with the familiar selection rules.  Only state
$|\uparrow\rangle$ has nonzero dipole matrix element with
$|\tau\rangle$, $d_{+}\mathbf{e}_+$, so that the transitions
$|1\rangle\rightarrow|\tau\rangle$ and
$|2\rangle\rightarrow|\tau\rangle$ have dipole matrix elements
equal to $d_{+}\cos\frac\psi 2\mathbf{e}_+$ and
$d_{+}\sin\frac\psi 2\mathbf{e}_+$ respectively. Then, for a
photon emitted along $\mathbf{n}_0=(\vartheta,\varphi)$, we find the couplings:
\begin{eqnarray}
g_{1\vartheta} &=& d_+ e^{i\varphi}\cos\vartheta \cos\frac\psi 2  \\
g_{1\varphi} &=& d_+ i e^{i\varphi} \cos\frac\psi 2\\
g_{2\vartheta} &=& d_+ e^{i\varphi}\cos\vartheta \sin\frac\psi 2 \\
g_{2\varphi} &=& d_+ i e^{i\varphi}\sin\frac\psi 2,
\end{eqnarray}
so that the determinant is always zero, independently of
$\mathbf{n}_0$. This means that the system in this configuration
will never be entangled with the polarization of the photon,
which, as we have seen, implies maximum SGC. The final state of
the $\Lambda$ system is always $|\uparrow \rangle$, unentangled.
Section~\ref{ppexp} gives an intuitive picture of this concept by
the vector representation of (the mean value of) the spin.

\subsection{Light hole trion in Voigt configuration}

The spin-photon entanglement can be also realized in a quantum dot
system by employing the light-hole trion state. The heavy and
light hole excitons are split by the breaking of the tetrahedral
symmetry of the bulk III-V compound. It might also be possible to
make the light hole states lower in energy than the heavy holes.
The magnetic field is pointing along the $x$ direction, so that
the lower levels are the two $S_x$ eigenstates, $|+\rangle$ and
$|-\rangle$. The optical pulses used are such that the light hole
trion polarized along the $+x$ direction is excited. The excited
state is a trion of a singlet pair of electrons and a light hole
which is in  the $m_j=\pm 1/2$ component of the $j=3/2$ state. The
trion can thus be characterized by the state
$|JM_J\rangle=|\frac{3}{2},\pm\frac{1}{2}\rangle$. We choose the
$M_J=\frac 1 2$ state as the excited state of  the $\Lambda$
system and denote it by $|\tau_l\rangle$.

The transitions $|\tau_l\rangle\rightarrow|+\rangle$ and
$|\tau_l\rangle\rightarrow|-\rangle$ involve a photon
linearly polarized along $x$ ($|X\rangle\equiv|\pi_x\rangle$) and one with elliptical
polarization ($-i|Y\rangle+2|Z\rangle\equiv|E_{yz}\rangle$), respectively \cite{lighthole_yabl}.
In particular, after $|\tau_l\rangle$
has decayed, the state of the system is from Eq.~(\ref{explicit}),
\begin{equation}
|\Upsilon\rangle = -\frac{1}{\sqrt{6}}[|X\rangle|-\rangle + (2|Z\rangle-i|Y\rangle)|+\rangle],\label{psilh}
\end{equation}
We assume a measurement which determines the propagation direction of the photon
$\mathbf{n}_0=(\vartheta,\varphi)$. Then the state becomes:
\begin{eqnarray}
|\Upsilon\rangle &=& \frac{-1}{\sqrt{2+3\sin^2\vartheta}}\big[
\cos\vartheta\cos\varphi|\vartheta\rangle|-\rangle \nonumber \\\nonumber \\
&&-(2\sin\vartheta+i\sin\varphi\cos\vartheta)|\vartheta\rangle|+\rangle \nonumber\\
&&-\sin\varphi|\varphi\rangle|-\rangle - i\cos\varphi|\varphi\rangle|+\rangle\big].
\label{psilh2}
\end{eqnarray}
Following the same procedure as in the trapped ion example, we
find that the condition for maximum entanglement is $\vartheta=0$;
the value of $D$ is then $0.5$, maximal entnaglement. SGC will
only occur when $D$ in Eq.~(\ref{maxent}) is less than 0.5 and it
will be maximum for propagation along $x$, which means that the
electron will be in the state $|+\rangle$. For all other values of
$\vartheta$ there will be both entanglement and SGC between the
two energy eigenstates when the photon is traced out. The
phenomena following the spontaneous radiative decay of this system
are indeed very similar to the trapped ion case. In the solid
state system there is no need to isolate a single dot in order to
observe SGC since all dots are oriented in the same direction.

\begin{figure}[t]
\begin{center}
\includegraphics[height=6.9cm,width=6.6cm, bb=90 110 310 340
,clip=true]{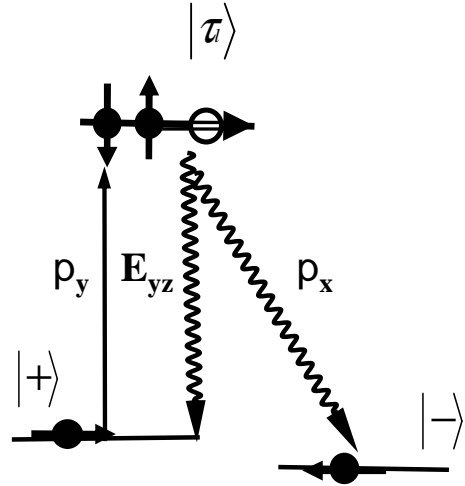}
\end{center}
\caption{The energy levels of the $\Lambda$ system consisting
of the two electron spin states (lower levels) and the light
hole trion polarized along the $+x$ direction. The solid line
represents the laser pulse, which propagates along $z$ and is
linearly polarized in the $y$ direction. The wavy lines denote
the spontaneously emitted photons from the transitions $|\tau_l\rangle\rightarrow|+\rangle$
and $|\tau_l\rangle\rightarrow|-\rangle$, which are elliptically polarized
in the $yz$ plane and linearly polarized along
$x$, respectively.}
\label{levels1}
\end{figure}

For quantum information processing, entanglement between
photon-polarization and spin has to be established in a quantum
dot. So isolating and addressing a single dot is required.
Experimentally, this requirement is arguably feasible
\cite{Steel_science}. The system should be initialized at state
$|+\rangle$ (or $|-\rangle$) and subsequently excited by $y$-(or
$x$-) polarized light, so that only  state $|\tau_l\rangle$ gets
excited. Other trion states, involving electrons in the triplet
state and/or heavy holes, have an energy separation from
$|\tau_l\rangle$ large enough compared to the pulse bandwidth and
so they can be safely ignored. Above we found that the state will
be maximally entangled when the spontaneously emitted photon
propagates along $z$. When the optical axis is along $z$, the
spontaneously emitted photon may be distinguished from the laser
photons by optical gating. As an alternative to the optical
gating, to minimize scattered light the detector may be placed
along $y$, i.e., at $(\vartheta,\varphi)=(\pi/2,\pi/2)$. The value
of $D$ is then 0.2, so that the entanglement will be significantly
less than that along the optical axis, but should be measurable.
The observation of the emitted photon and the measurement of its
polarization can be made as in Ref.~\cite{monroe}. By use of the
pump-probe technique, the state of the spin will also be measured
to show the correlation with the polarization of the photon.

To overcome the probabilistic nature of the entanglement (as
projection is needed) and to improve the quantum efficiency
degraded by the scattering problem, cavities and waveguides may be
employed to enhance and select desired photon emission processes
\cite{ReadWrite,qinterface}.

\section{Pump-probe experiment for SGC detection in a quantum dot} \label{ppexp}

In this section we provide a theoretical analysis for the
pump-probe experiment which explicitly demonstrated SGC
\cite{Gurudev}. The $\Lambda$ system is the heavy-hole trion
system introduced above. We present a treatment based on the idea
that SGC may be viewed as a decay to one bright state which is a
superposition of the eigenstates. The vector character of the mean
value of the spin, which also helps develop intuition for the SGC
effect, is  employed and in fact it anticipates some of the
theoretical results of the pump-probe measurements calculated by
perturbative solution of the density matrix in the remainder of
this section.

\subsection{Geometrical picture of SGC}
\label{intuitive}

As shown by Bloch \cite{bloch} and Feynman {\it et al}
\cite{Feynman_spin}, an ensemble of two-level systems can be
described by a rotating vector. This picture provides an intuitive
understanding of the spin coherence generated by the optical
excitation and spontaneous decay of the trion states. For
simplicity, we will assume the short-pulse limit in this section.

 Regardless of the presence or absence of the magnetic field,
there is freedom in the choice of the quantization direction, and
it is convenient in this case to choose the spin eigenstates
quantized in the growth ($z$) direction, $|\uparrow\rangle$ and
$|\downarrow\rangle$. The two trion states $|\tau\rangle$ and
$|\bar{\tau}\rangle$ have $J= 3/2$ and $z$-component $M=+3/2$ and
$M=-3/2$, respectively. The selection rules are such that a photon
with helicity $\pm 1$ ($\sigma{\pm}$ circular polarization)
excites the electron $|\uparrow\rangle$ or $|\downarrow\rangle$ to
the trion states $|\tau\rangle$ or $|\bar{\tau}\rangle$,
respectively. We will consider a $\sigma+$ polarized pump, which
excites spin-up electrons to the trion state $|\tau\rangle$,
leaving the electron spin-polarized in the $-z$ direction. Due to
the selection rules, the trion state can only relax back to the
spin up state by emitting a $\sigma+$ polarized photon, and after
recombination, the electron remains unpolarized.

\begin{figure}[t]
\begin{center}
\includegraphics[height=6.75cm,width=8.1cm, bb=150 525 420 750,clip=true]{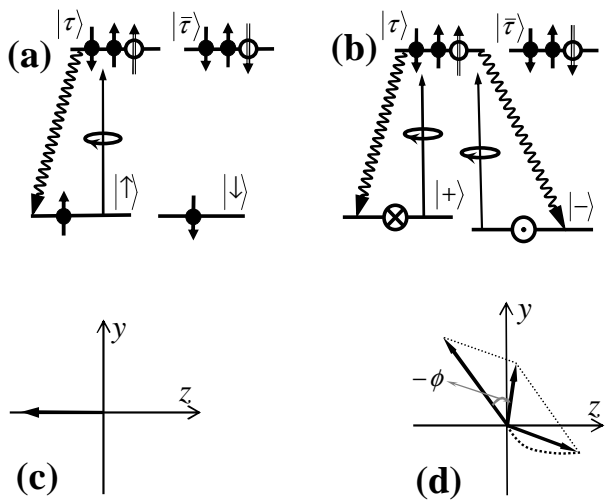}
\end{center}
\caption{(a) and (b) are the energy diagrams and possible
electron-trion transitions caused by $\sigma+$-polarized photons
with the electron spin quantized in $z$ and $x$ directions,
respectively. (c) plots the Raman coherence generated by the pump
pulse, and (d) schematically depicts interference between the
Raman coherence and the spin coherence generated by spontaneous
emission, under a magnetic field applied along the $x$ direction.
\label{levels}}
\end{figure}

Now let us consider a strong magnetic field, applied at
$\psi=\pi/2$ with respect to the optical axis,
$\bm{B}=B\mathbf{e}_x$. In this so-called Voigt configuration, the
Zeeman states $|\pm\rangle\equiv
(|\uparrow\rangle\pm|\downarrow\rangle)/\sqrt{2}$ are quantized in
the $x$-direction and are energy eigenstates with energies $\pm
\omega_{\rm L}$, respectively, while the trion states can still be
assumed quantized in the $z$-direction [see Fig.~\ref{levels}
(b)]. Note that the low-lying states $|1\rangle, |2\rangle$ in
foregoing sections are now denoted by the spin states, $|+\rangle,
|-\rangle$.
 In the short-pulse limit, the pulse spectrum is much broader
than the spin splitting or, equivalently, the pulse duration is
much shorter than the spin precession period, so the excitation
process is virtually unaffected by the magnetic field: the
$\sigma+$ polarized pump excites the $|\uparrow\rangle$ electron
to the trion state $|\tau\rangle$, leaving the electrons
spin-polarized in the $-z$ direction, as in the zero-field case
[See Fig.~\ref{levels} (c)]. The pulse generates coherence between
the two eigenstates $|+\rangle$ and $|-\rangle$, which is the
conventional Raman coherence \cite{Raman_Coherence} generated by a
pulse with a spectrum broad enough to cover both the
near-degenerate transitions. The spin precesses in the magnetic
field normal to the plane of precession with frequency
$\omega_{\rm L}/\pi$. In other words, the state oscillates between
the spin up and down states. The Raman coherence can be determined
by the excitation-induced change of the population in the spin
state $|\uparrow\rangle$,
\begin{equation}
\rho^{R}_{\uparrow\uparrow}(t)=-{\rho_{\tau\tau}\over
2}\left[1+\cos\left(2\omega_{\rm L}t\right)e^{-\gamma_2 t}\right],
\end{equation}
where $\rho_{\tau\tau}$ is the population of the trion state
immediately after the excitation pulse, and $\gamma_2$ is the
damping rate of the spin polarization (due to spin dephasing
and inhomogeneous broadening).

On the other hand, when the system is in the trion state
$|\tau\rangle$, the trion will relax by emitting a
$\sigma+$-polarized photon, leaving an electron spin-polarized in
the $+z$ direction, i.e., generating coherence between the two
spin eigenstates (SGC). The trion decay can be treated as a
stochastic quantum jump process with the jump rate $2\Gamma$.
After the quantum jump, the evolution of the system can be
described by a spin vector rotating under the transverse magnetic
field. Thus, the spin polarization generated by the spontaneous
emission during $[t',t'+dt']$ can be determined by
\begin{eqnarray}
d\rho^{SGC}_{\uparrow\uparrow}(t,t')&=&\frac{{\rho_{\tau\tau}}e^{-2\Gamma
t'}2\Gamma dt'}{2}\times \nonumber \\
&&{\left[1+\cos\left(2\omega_{\rm
L}(t-t')\right)e^{-\gamma_2 (t-t')}\right]}. \label{drhosgc}
\end{eqnarray}
The precessing spin vector is deformed by the accumulation of
increments through the optical decay into a spiral curve [see Fig.
\ref{levels} (d)].
The accumulated spin polarization due to the
spontaneous emission is
\begin{eqnarray}
\rho^{SGC}_{\uparrow\uparrow}(t)=\int_0^{t}
d\rho^{SGC}_{\uparrow\uparrow}(t, t')= \qquad\qquad\qquad\qquad\qquad&&\nonumber \\
{\rho_{\tau\tau}\over
2}\Re\!\left[\!1\!-\!e^{-2\Gamma
t}\!+\!\frac{2\Gamma}{2\Gamma-\gamma_2-i2\omega_{\rm L}}
\!\left(\!e^{-i2\omega_{\rm L} t-\gamma_2 t}\!-\!e^{-2\Gamma t}\right)
\!\right]. \nonumber\\
\label{rhosgc}
\end{eqnarray}
For an initially unpolarized system, the total spin polarization
in the $z$ direction after the action of the pump and the
recombination process is given by
\begin{eqnarray}
\rho_{\uparrow\uparrow}^{(2)}=
\left[\rho^R_{\uparrow\uparrow}+\rho^{SGC}_{\uparrow\uparrow}\right]\qquad\qquad\qquad\qquad\qquad\qquad\nonumber\\
= -{\rho_{\tau\tau}\over 2}\left[\left(1+a_{\Gamma}\right)
e^{-2\Gamma t}+a_0\cos\left(2\omega_{\rm L} t-\phi\right)
e^{-\gamma_2 t}\right] ,\label{beat0}
\end{eqnarray}
where
\begin{eqnarray}
 a_{\Gamma}&\equiv& {2\Gamma\left(2\Gamma-\gamma_2\right)\over
 \left(2\Gamma-\gamma_2\right)^2+4\omega_{\rm L}^2}, \\
 a_{0}&\equiv& \sqrt{\gamma_2^2+4\omega_{\rm L}^2\over
\left(2\Gamma-\gamma_2\right)^2+4\omega_{\rm L}^2}, \\
 \phi &\equiv& -\arctan {2\Gamma-\gamma_2\over 2\omega_{\rm L}}
-\arctan{\gamma_2\over 2\omega_{\rm L}}.
\end{eqnarray}

As shown in Fig. \ref{levels} (d), SGC induces a phase shift of the
spin coherence as compared to the Raman coherence.  Note also the different
amplitudes of the Bloch vectors in the case with and without SGC.
We can see that if the recombination is much faster than the spin
precession under the magnetic field, i.e., $\Gamma\gg\omega_{\rm
L}$, SGC actually cancels the Raman coherence. This is not
surprising since such a limit simply corresponds to the zero-field
case. In the strong field limit where $\omega_{\rm L}\gg\Gamma$,
the spin precession will average SGC to zero, which corresponding to
the two-pathway decay discussed in Sec. \ref{unified}. From Eq.~(\ref{drhosgc})
it can be seen that at any specific time the trion relaxes to state
$|\uparrow\rangle$, so, as shown in Sec. \ref{unified}, a time-selective
measurement can recover the SGC from the incoherent two-pathway decay. Without
such a projection, as the spin coherence generated at different time has different
phaseshift, the time averaging [see Eq. (\ref{rhosgc})] leads to the vanishing of
the SGC.

In a pump-probe experiment, what is measured is the differential
transmission signal (DTS), i.e., the difference between the probe
transmission with and without the pump pulse. In the same-circular
polarization (SCP) pump-probe configuration, the probe measures
the change in the population difference created by the pump,
$\rho_{\tau\tau}-\rho^{(2)}_{\uparrow\uparrow}$. Hence, the DTS is
given by
\begin{eqnarray}
 \Delta T^{\rm SCP}  \propto  \left(3+
a_{\Gamma}\right)e^{-2\Gamma {t_{\rm d}}}  + a_0
\cos\left(2\omega_{\rm L} {t_{\rm d}} -\phi\right),
\end{eqnarray}
where ${t_{\rm d}}$ is the delay time between the pump and probe
pulses. The DTS reveals the spin beatings and the SGC effect
manifests itself in the dependence of the beat amplitude and phase
shift on the strength of the magnetic field.

The pump-probe experiment can also be done in the opposite
circular polarization (OCP) configuration. The probe measures the
change of population of the spin down state $|\downarrow\rangle$.
The DTS in this case is proportional to,
$-\rho^{(2)}_{\downarrow\downarrow}=\rho^{(2)}_{\uparrow\uparrow}+\rho_{\tau\tau}$,
i.e.,
\begin{eqnarray}
 \Delta T^{\rm OCP}\propto \left(1- a_{\Gamma}\right)e^{-2\Gamma
{t_{\rm d}}}-a_0 \cos\left(2\omega_{\rm L} {t_{\rm d}}
-\phi\right).
\end{eqnarray}
The spin beat has the opposite sign to the SCP case.

Similar analysis shows that if either the pump or the probe pulse
is linearly polarized there will be no spin beat in the DTS.

\subsection{Perturbative solution of the probe signal} \label{Master}

The optical field of the pump and probe pulses can be written as
\begin{eqnarray}
{\mathbf E}(t) &=&  \left({\mathbf e}_+E_{{1}+}+{\mathbf
e}_-E_{{1}-}\right)\chi_{{1}}(t)e^{-i\Omega_{1} t} \nonumber
\\
&+&\left({\mathbf e}_+E_{{2}+}+{\mathbf
e}_-E_{{2}-}\right)\chi_{{2}}(t-{t_{\rm d}})e^{-i\Omega_{2}
(t-{t_{\rm d}})},
\end{eqnarray}
where the subscripts ${1}$ and ${2}$ denote the pump and probe
pulses, respectively, and ${\mathbf e}_{\pm}$ are the unit vectors
of the $\sigma{\pm}$-polarizations. The dipole operator is
$${\hat {\mathbf d}}=d\Big({\mathbf e}_+|\tau\rangle\langle
\mp| \pm {\mathbf e}_-|\bar{\tau}\rangle\langle \pm|\Big) +{\rm
h.c.}.$$ Thus, in the rotating wave approximation, the Hamiltonian
in the basis $\{|-\rangle,|+\rangle, |\tau\rangle,
|\bar{\tau}\rangle\}$ can be written in matrix form as
\begin{equation}
H=
\left[\begin{array}{cccc}-\omega_{\rm L}& 0 & -d^*E^*_+(t) &  -d^*E^*_-(t)\\
0 & \omega_{\rm L} & -d^*E^*_+(t) & +d^*E^*_-(t)\\
-d E_+(t) & -dE_+(t)& \epsilon_g &0 \\
-d E_-(t) & +dE_-(t)& 0 & \epsilon_g
\end{array}\right] ,
\end{equation}
where $\epsilon_g$ is the energy of the trion states, and
$\gamma_1$, $\gamma_2$, and $2\Gamma$ denoting the spin-flip rate,
the spin depolarizing rate, and the trion decay rate,
respectively. The explicit equations for each element of the
density matrix are
\begin{eqnarray}
\dot{\rho}_{\tau,+}&=& i[\rho,H]_{\tau,+} - \Gamma\rho_{\tau,+}, \\
\dot{\rho}_{\tau,-}&=&i[\rho,H]_{\tau,-} - \Gamma\rho_{\tau,-}, \\
\dot{\rho}_{+,+}&=&i[\rho,H]_{+,+} - \gamma_1\rho_{+,+} +
\Gamma\left(\rho_{\tau\tau}+\rho_{\bar{\tau} , \bar{\tau}}\right), \\
\dot{\rho}_{-,-}&=&i[\rho,H]_{-,-} + \gamma_1\rho_{+,+} +
\Gamma\left(\rho_{\tau\tau}+\rho_{\bar{\tau} \bar{\tau}}\right), \\
\dot{\rho}_{+,-}&=&i[\rho,H]_{+,-} -\gamma_2\rho_{+,-}
 +\Gamma_{\rm c}\left(\rho_{\tau\tau}-\rho_{\bar{\tau}, \bar{\tau}}\right), \\
\dot{\rho}_{\tau\tau}&=&i[\rho,H]_{\tau\tau} - 2\Gamma\rho_{\tau\tau}, \\
\dot{\rho}_{\bar{\tau},{t}}&=&i[\rho,H]_{\bar{\tau},{t}} - 2\Gamma\rho_{\bar{\tau},{t}}, \\
\dot{\rho}_{\bar{\tau}\bar{\tau}}&=&i[\rho,H]_{\bar{\tau}\bar{\tau}}
- 2\Gamma\rho_{\bar{\tau}\bar{\tau}}.
\end{eqnarray}
The Markov-Born approximation for the system-photon has been
employed. The term representing the spontaneously generated spin
coherence due to the trion recombination is indicated by the
suffix $c$; $\Gamma_{\rm c}$ should be equal to $\Gamma$. However,
we singled out the SGC term so that $\Gamma_{\rm c}$ can be
artificially set to zero for a theoretical comparison between the
results with and without the SGC effect.

In the pump-probe experiment, the DTS corresponds to the
third-order optical response. The absorption of the probe pulse is
proportional to the work $W$ done by the probe pulse, and the DTS
is \cite{bloem}
\begin{eqnarray}
\Delta T &\propto & -W^{(3)}= -2\Re\int \dot{\mathbf
P}^{(3)}(t)\cdot {\mathbf E}_{2}^{*}(t-{t_{\rm d}}) \nonumber \\
& \approx & -2 \Omega_{2} \Im\int
\tilde{\mathbf{P}}^{(3)}(\omega+\Omega_{2})\cdot \tilde{\mathbf
E}^*_{2}(\omega+\Omega_{2}) \frac{d\omega}{2\pi}. \label{DTS}
\end{eqnarray}
 The third-order optical
polarization of the system can be calculated directly by expanding
the density matrix according to the order of the optical
perturbation
\begin{eqnarray}
{\mathbf P}^{(3)}={\mathbf e}_+
d\left[\rho^{(3)}_{\tau,-}+\rho^{(3)}_{\tau,+}\right]+{\mathbf
e}_-
d\left[\rho^{(3)}_{\bar{\tau},-}-\rho^{(3)}_{\bar{\tau},+}\right],
\end{eqnarray}
 Thus, given the $\sigma+$-polarized pump pulse, the third-order polarization
in the SCP and OCP cases can be respectively calculated as
\cite{bloem}
\begin{eqnarray}
{\mathbf P}_{\rm SCP}^{(3)}(t)={\mathbf e}_+d\left[\rho^{(3)}_{\tau,-}(t)+\rho^{(3)}_{\tau,+}(t)\right], \\
{\mathbf P}_{\rm OCP}^{(3)}(t)={\mathbf
e}_-d\left[\rho^{(3)}_{\bar{\tau},-}(t)-\rho^{(3)}_{\bar{\tau},+}(t)\right].
\end{eqnarray}

\subsection{Analytical results} \label{analytical}

The density matrix can be calculated straightforwardly order by
order with respect to the pulse. Taking the initial state of the
system to be the equilibrium state
$\rho^{(0)}=\rho^{(0)}_+|+\rangle\langle
+|+\rho^{(0)}_-|-\rangle\langle -|$. The result for the
second-order spin coherence due to the pump pulse ${\mathbf
E}_{1}(t)$ is:
\begin{widetext}
\begin{eqnarray}
&& {\tilde \rho}^{(2)}_{+-}(\omega)=+
   X_{1}
    {{\rho^{(0)}_{-}}\over
   \omega-2\omega_{\rm L}+i\gamma_2}
   \int^{+\infty}_{-\infty}{\chi^*_{1}(\omega'-\omega)\chi_{1}(\omega')\over
   \omega'-\Delta_{1}-\omega_{\rm L}+i\Gamma}{d\omega'\over
   2\pi}
   \nonumber \\ &&\phantom{ {\tilde \rho}^{(2)}_{+,-}(\omega)=}
   -
   X_{1}
    {{\rho^{(0)}_{+}}\over
   \omega-2\omega_{\rm L}+i\gamma_2}
   \int^{+\infty}_{-\infty}{\chi_{1}(\omega'+\omega)\chi^*_{1}(\omega')\over
   \omega'-\Delta_{1}+\omega_{\rm L}-i\Gamma}{d\omega'\over
   2\pi} \nonumber \\
&&\phantom{ {\tilde \rho}^{(2)}_{+,-}(\omega)=}+
  X_{1}
    {i\Gamma_{\rm c}{\rho^{(0)}_{\pm}}\over
   (\omega-2\omega_{\rm L}+i\gamma_2)(\omega+i2\Gamma)}
   \int^{+\infty}_{-\infty}
   {\chi_{1}(\omega'+\omega)\chi^*_{1}(\omega')\over
   \omega'-\Delta_{1}\pm\omega_{\rm L}-i\Gamma}
   {d\omega'\over
   2\pi} \nonumber \\
&&\phantom{ {\tilde \rho}^{(2)}_{+,-}(\omega)=}-
  X_{1}
    {i\Gamma_{\rm c}{\rho^{(0)}_{\pm}}\over
   (\omega-2\omega_{\rm L}+i\gamma_2)(\omega+i2\Gamma)}
   \int^{+\infty}_{-\infty}
   {\chi^*_{1}(\omega'-\omega)\chi_{1}(\omega')\over
   \omega'-\Delta_{1}\pm\omega_{\rm L}+i\Gamma}{d\omega'\over
   2\pi}, \label{coherence}
\end{eqnarray}
\end{widetext}
where $\Delta_{1}\equiv \epsilon_g-\Omega_{1}$ is the detuning, and
$X_{1}\equiv \left|dE_{{1}+}\right|^2-\left|dE_{{1}-}\right|^2$ is
the circular degree of the pulse polarization. In the equation
above, the first two terms correspond to the Raman coherence
generated by the pump excitation \cite{Raman_Coherence}, and the
last two terms represent the spontaneously generated coherence.
Obviously, for a linearly polarized pump, $X_{1}=0$, no spin
coherence is generated either by excitation or by recombination,
so there will be no spin beats in DTS.

In the short-pulse limit, the spin coherence after the pump and
recombination can be approximately expressed as
\begin{eqnarray}
\rho^{(2)}_{+,-}(t) &\approx&
X_{1}\left|\chi_{1}(\Delta_{1})\right|^2 \left(\frac{\Gamma_{\rm
c}}{2\Gamma-\gamma_2-2i\omega_{\rm L}}-{1\over 2}\right) \nonumber
\\ && \times e^{-i(2\omega_{\rm L}-i\gamma_2)(t-t_{1})}. \label{short}
\end{eqnarray}
This formula can be directly compared to the result obtained by
the intuitive picture in Sec.~\ref{intuitive}. The physical
meaning of the two terms in Eq.~(\ref{short}) is transparent: the
first term is SGC, whose amplitude and phase shift depend on the
ratio of the recombination rate to the Zeeman splitting, and the
second term is just the optically pumped Raman coherence which in
the short pulse limit is independent of the Zeeman splitting.

Having obtained the second-order results, we can readily derive
the third-order density matrix and, in turn, the DTS can be
calculated by use of Eq. (\ref{DTS}). In general, the DTS can be
expressed as
\begin{equation}
\Delta T\propto {A} \cos(2\omega_{\rm L}{t_{\rm d}} -
\phi)e^{-\gamma_2{t_{\rm d}}} + {B} e^{-2\Gamma{t_{\rm
d}}}+{C}e^{-\gamma_1 {t_{\rm d}}}  , \label{old13}
\end{equation}
and the spin coherence amplitude ${A}$ and phase shift $\phi$, the
Pauli blocking amplitude ${B}$, and the spin non-equilibrium
population ${C}$ can all be numerically calculated and, in the
short-pulse limit, can also be analytically derived as
\begin{eqnarray}
{A} &\approx&\left|\chi_{1}\left(\Delta_{1}\right)\right|^2
 \left|\chi_{2}\left(\Delta_{2}\right)\right|^2  X_{1}X_{2} \nonumber \\
 && \times
\sqrt{\frac{{\gamma_2}^2+4\omega^2_{\rm L}}{(2\Gamma_{\rm
c}-{\gamma_2})^2+4\omega^2_{\rm L}}},
\\
\phi&\approx& -\arctan\left(\frac{2\Gamma_{\rm
c}-\gamma_2}{2\omega_{\rm
L}}\right)-\arctan\left(\frac{\gamma_2}{2\omega_{\rm L}}\right),
\label{old16}\\
{B}&\approx& \left|\chi_{1}\left(\Delta_{1}\right)\right|^2
 \left|\chi_{2}\left(\Delta_{2}\right)\right|^2
\Big[I_{1}I_{2}+2I_{{1}+}I_{{2}+}
 \nonumber\\ && \ \
  +2I_{{1}-}I_{{2}-} +X_{{1}}X_{{2}}\frac{2\Gamma_{\rm c}
(2\Gamma-\gamma_2)}{(2\Gamma-{\gamma_2})^2+4\omega^2_{\rm L}}\Big], \\
{C} &\approx& 0,
\end{eqnarray}
where $\Delta_{2}\equiv \epsilon_g-\Omega_{2}$ is the detuning,
$I_{j\pm}\equiv \big|dE_{j\pm}\big|^2$, $I_{j}\equiv
I_{j+}+I_{j-}$, and $X_j\equiv I_{j+}-I_{j-}$  ($j={1}$ or ${2}$).
Thus, the short-pulse approximation yields expressions identical
to the ones obtained from the intuitive picture in Section
\ref{intuitive}. Several conclusions can be immediately drawn from
the short-pulse approximation:
(1) the SCP and OCP signals reveal
beats with the same amplitude and opposite signs; (2) no spin beat
can be observed when either of the pulses is linearly polarized
($X_{1}=0$ or $X_{2}=0$); (3) due to the SGC effect, the beat
amplitude increases with increasing Zeeman splitting until it
saturates at the value it would have in the absence of the SGC
effect; the phase shift increases from $-\pi/2$, saturating at 0.
The SGC effect is negligible when the Zeeman splitting is large
compared to the trion decay rate $\Gamma$ because the rapid
oscillation averages the effect of SGC to zero.

\subsection{Numerical results} \label{numerical}

\begin{figure}[t]
\begin{center}
\includegraphics[height=6.79cm,width=7.5cm, bb=45 220 570 695, clip=true]{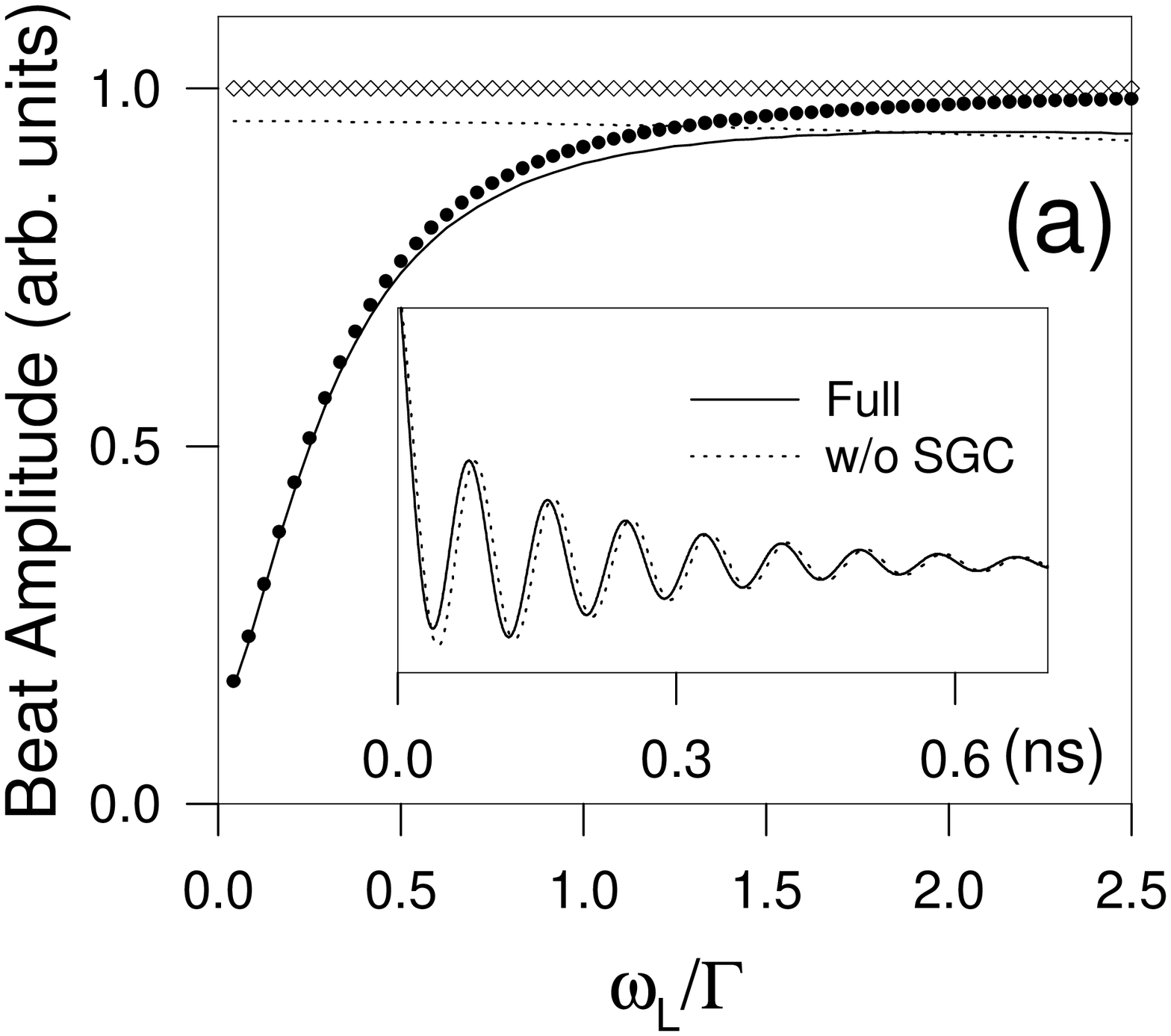} \vskip .5cm
\includegraphics[height=6.72cm,width=7.5cm, bb=40 220 570 695, clip=true]{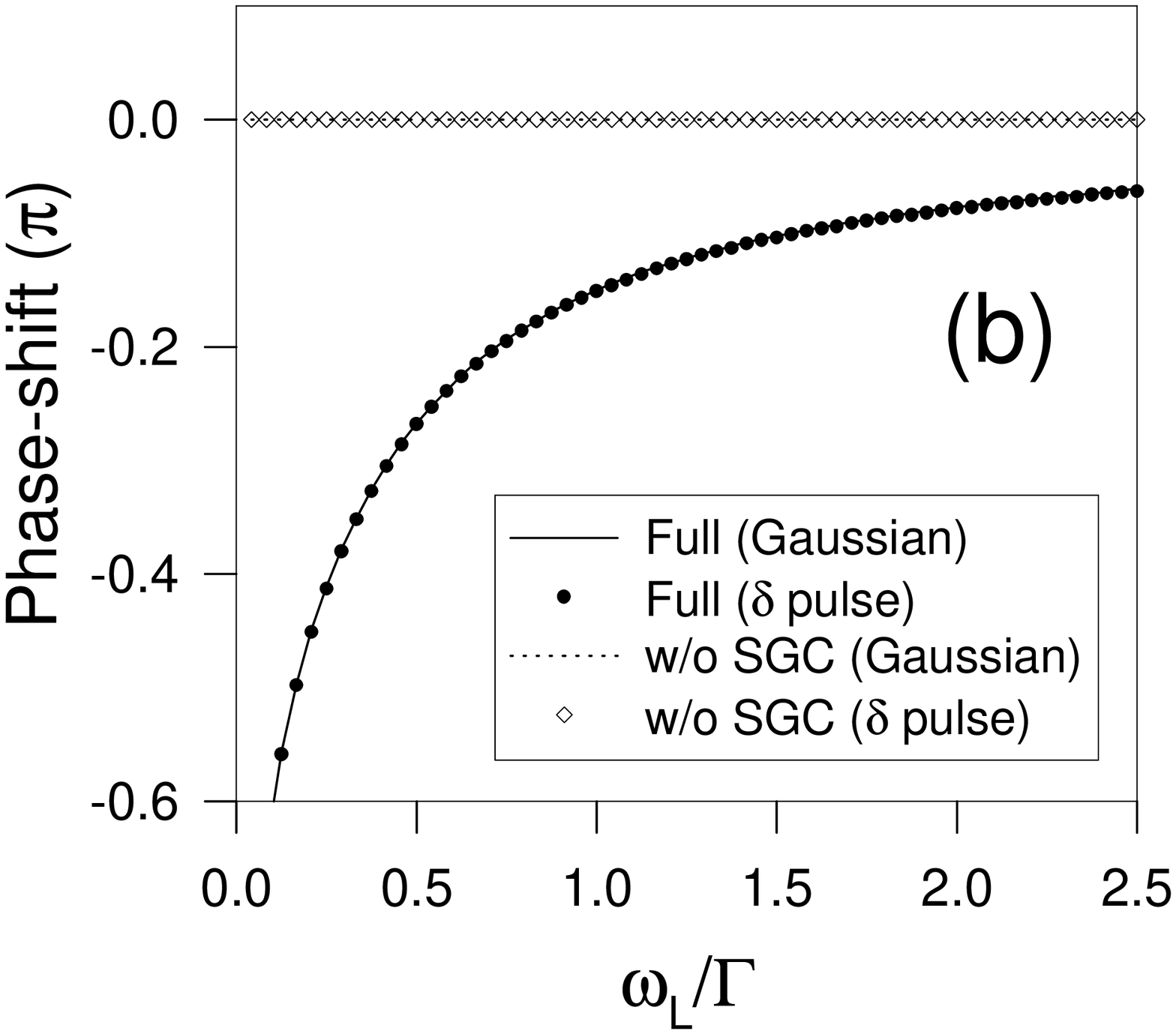}
\end{center}
\caption{(a) The amplitude and (b) the phase shift of the spin beat (shown
in the insert)
as functions of the Zeeman splitting in units of the trion
state width, $\Gamma$. The fulled-circle and solid lines include the SGC
effect, calculated with and without the short-pulse approximation,
respectively. The diamond and dotted lines are the results
without the SGC effect, calculated with and without the
short-pulse approximation, respectively. \label{fig:wide2}}
\end{figure}

In the numerical simulations, we take the pump and the probe
envelopes to be gaussian $\chi_{1}(t)=\exp\left(-\eta_{1}^2t^2/2\right)$ and
$\chi_{2}(t-{t_{\rm d}})=\exp\left(-\eta_{2}^2(t-{t_{\rm
d}})^2/2\right)$, and we assume that they have no temporal
overlap, i.e. the delay time ${t_{\rm d}}$ is much larger
than the pulse duration $\eta^{-1}_j$ ($j={1},{2}$), and the pulse
bandwidth $\eta_j$ is greater than the relaxation rates
$\gamma_1$,  $\gamma_2$, and $\Gamma$. All these assumptions are
well satisfied in the experiment \cite{Gurudev}.  Taken from the
experiment \cite{Gurudev}, the relaxation rates used are
$\gamma_1=0$, $\gamma_2=3\ \mu$eV, and $\Gamma=\Gamma_{\rm c}=12\
\mu$eV, and $\eta_{1}=\eta_{2}=0.5$ meV.

To minimized  the effect of the background noise \cite{Gurudev},
the measured data of DTS  are presented as  the difference between
the SCP and OCP. We follow the same practice in presenting the
theoretical results in Fig.~\ref{fig:wide2}. In comparison with
the results without the SGC effect (dashed line), the full
theoretical results show the phase shift of the spin beat in the
DTS.

 In Fig.~\ref{fig:wide2}, the amplitude and the phase shift are
plotted against the Zeeman splitting $2\omega_{\rm L}$, which is
proportional to the magnetic field.  The SGC effect is evident
through the field dependence of the amplitude and phase shift of
the spin beat. When the SGC effect is artificially switched off
(by setting $\Gamma_{\rm c}=0$), the beat is independent of the
magnetic field strength as long as the pulse spectrum is much
broader than the Zeeman splitting. In the weak magnetic field
limit, the spin coherence is strongly suppressed due to the
destructive interference between the conventional Raman coherence
and SGC; the phase shift then is about $-\pi/2$. In the strong
magnetic field limit, as SGC is averaged to zero due to the rapid
Larmor precession, the beat features approach those calculated
without SGC. The theoretical predictions of the SGC effect on the
pump-probe signals are in good agreement with the experimental
results \cite{Gurudev}.

\section{Conclusions}
\label{conclusions} In this work, we have developed a theory to
unite the different effects emerging from the spontaneous emission
of a photon from a $\Lambda$ system. We have taken the viewpoint
that spontaneous emission is a unitary process when a sufficiently
large quantum system is defined so as to be considered closed.
Then the final state of the whole system, which is a pure state,
can be projected in different ways. These projections can be
thought of as measurements on one of the constituent parts and
give rise to different phenomena: entanglement, spontaneously
generated coherence and two-pathway decay. We have also presented
a set of conditions on the symmetry of a system which determine if
there is SGC. Examples of specific atomic and solid-state systems
have been employed to illustrate our theory. We have sketched the
theory underlying the experiment in which SGC was observed
\cite{Gurudev} and we have proposed an experiment on the same
system to exhibit the entanglement between the electron spin and
the polarization of the spontaneously emitted photon in a quantum
dot in parallel to the atom case \cite{monroe_atom}.

\begin{acknowledgments}
This work was supported by ARDA/ARO under grant W911NF-04-1-0235
and a Graduate Fellowship DAAD19-02-10183. S.E.E also acknowledges
a graduate fellowship from the Alexander S. Onassis Public Benefit
Foundation.
\end{acknowledgments}


\begin{thebibliography}{29}
\expandafter\ifx\csname
natexlab\endcsname\relax\def\natexlab#1{#1}\fi
\expandafter\ifx\csname bibnamefont\endcsname\relax
  \def\bibnamefont#1{#1}\fi
\expandafter\ifx\csname bibfnamefont\endcsname\relax
  \def\bibfnamefont#1{#1}\fi
\expandafter\ifx\csname citenamefont\endcsname\relax
  \def\citenamefont#1{#1}\fi
\expandafter\ifx\csname url\endcsname\relax
  \def\url#1{\texttt{#1}}\fi
\expandafter\ifx\csname
urlprefix\endcsname\relax\def\urlprefix{URL }\fi
\providecommand{\bibinfo}[2]{#2}
\providecommand{\eprint}[2][]{\url{#2}}

\bibitem[{\citenamefont{Harris et~al.}(1990)\citenamefont{Harris, Field, and
  Imamo\u{g}lu}}]{EIT}
\bibinfo{author}{\bibfnamefont{S.~E.} \bibnamefont{Harris}},
  \bibinfo{author}{\bibfnamefont{J.~E.} \bibnamefont{Field}}, \bibnamefont{and}
  \bibinfo{author}{\bibfnamefont{A.}~\bibnamefont{Imamo\u{g}lu}},
  \bibinfo{journal}{Phys. Rev. Lett.} \textbf{\bibinfo{volume}{64}},
  \bibinfo{pages}{1107} (\bibinfo{year}{1990}).

\bibitem[{\citenamefont{Harris}(1989)}]{LaserNoInv}
\bibinfo{author}{\bibfnamefont{S.~E.} \bibnamefont{Harris}},
  \bibinfo{journal}{Phys. Rev. Lett.} \textbf{\bibinfo{volume}{62}},
  \bibinfo{pages}{1033} (\bibinfo{year}{1989}).

\bibitem[{\citenamefont{Imamo\u{g}lu et~al.}(1999)\citenamefont{Imamo\u{g}lu,
  Awschalom, Burkard, DiVincenzo, Loss, Sherwin, and
  Small}}]{Imamoglu_CQED_Spin}
\bibinfo{author}{\bibfnamefont{A.}~\bibnamefont{Imamo\u{g}lu}},
  \bibinfo{author}{\bibfnamefont{D.~D.} \bibnamefont{Awschalom}},
  \bibinfo{author}{\bibfnamefont{G.}~\bibnamefont{Burkard}},
  \bibinfo{author}{\bibfnamefont{D.~P.} \bibnamefont{DiVincenzo}},
  \bibinfo{author}{\bibfnamefont{D.}~\bibnamefont{Loss}},
  \bibinfo{author}{\bibfnamefont{M.}~\bibnamefont{Sherwin}}, \bibnamefont{and}
  \bibinfo{author}{\bibfnamefont{A.}~\bibnamefont{Small}},
  \bibinfo{journal}{Phys. Rev. Lett.} \textbf{\bibinfo{volume}{83}},
  \bibinfo{pages}{4204} (\bibinfo{year}{1999}).

\bibitem[{\citenamefont{Chen et~al.}(2004)\citenamefont{Chen, Piermarocchi,
  Sham, Gammon, and Steel}}]{OpticalControl_Sham}
\bibinfo{author}{\bibfnamefont{P.}~\bibnamefont{Chen}},
  \bibinfo{author}{\bibfnamefont{C.}~\bibnamefont{Piermarocchi}},
  \bibinfo{author}{\bibfnamefont{L.~J.} \bibnamefont{Sham}},
  \bibinfo{author}{\bibfnamefont{D.}~\bibnamefont{Gammon}}, \bibnamefont{and}
  \bibinfo{author}{\bibfnamefont{D.~G.} \bibnamefont{Steel}},
  \bibinfo{journal}{Phys. Rev. B} \textbf{\bibinfo{volume}{69}},
  \bibinfo{pages}{075320} (\bibinfo{year}{2004}).

\bibitem[{\citenamefont{Monroe et~al.}(1995)\citenamefont{Monroe, Meekhof,
  King, Itano, and Wineland}}]{monroe_atom}
\bibinfo{author}{\bibfnamefont{C.}~\bibnamefont{Monroe}},
  \bibinfo{author}{\bibfnamefont{D.~M.} \bibnamefont{Meekhof}},
  \bibinfo{author}{\bibfnamefont{B.~E.} \bibnamefont{King}},
  \bibinfo{author}{\bibfnamefont{W.~M.} \bibnamefont{Itano}}, \bibnamefont{and}
  \bibinfo{author}{\bibfnamefont{D.~J.} \bibnamefont{Wineland}},
  \bibinfo{journal}{Phys.\ Rev. Lett.} \textbf{\bibinfo{volume}{75}},
  \bibinfo{pages}{4714} (\bibinfo{year}{1995}).

\bibitem[{\citenamefont{Brennen et~al.}(1999)\citenamefont{Brennen, Caves,
  Jessen, and Deutsch}}]{OpticallatticeQC}
\bibinfo{author}{\bibfnamefont{G.~K.} \bibnamefont{Brennen}},
  \bibinfo{author}{\bibfnamefont{C.~M.} \bibnamefont{Caves}},
  \bibinfo{author}{\bibfnamefont{P.~S.} \bibnamefont{Jessen}},
  \bibnamefont{and} \bibinfo{author}{\bibfnamefont{I.~H.}
  \bibnamefont{Deutsch}}, \bibinfo{journal}{Phys.\ Rev. Lett.}
  \textbf{\bibinfo{volume}{82}}, \bibinfo{pages}{1060} (\bibinfo{year}{1999}).

\bibitem[{\citenamefont{Liu et~al.}(2004)\citenamefont{Liu, Yao, and
  Sham}}]{ReadWrite}
\bibinfo{author}{\bibfnamefont{R.~B.} \bibnamefont{Liu}},
  \bibinfo{author}{\bibfnamefont{W.}~\bibnamefont{Yao}}, \bibnamefont{and}
  \bibinfo{author}{\bibfnamefont{L.~J.} \bibnamefont{Sham}},
  \bibinfo{journal}{cond-mat/0408148}  (\bibinfo{year}{2004}).

\bibitem[{\citenamefont{Blinov et~al.}(2004)\citenamefont{Blinov, Moehring,
  Duan, and Monroe}}]{monroe}
\bibinfo{author}{\bibfnamefont{B.~B.} \bibnamefont{Blinov}},
  \bibinfo{author}{\bibfnamefont{D.~L.} \bibnamefont{Moehring}},
  \bibinfo{author}{\bibfnamefont{L.-M.} \bibnamefont{Duan}}, \bibnamefont{and}
  \bibinfo{author}{\bibfnamefont{C.}~\bibnamefont{Monroe}},
  \bibinfo{journal}{Nature (London)} \textbf{\bibinfo{volume}{428}},
  \bibinfo{pages}{153} (\bibinfo{year}{2004}).

\bibitem[{\citenamefont{Cohen-Tannoudji
  et~al.}(1992)\citenamefont{Cohen-Tannoudji, Dupont-Roc, and
  Grynberg}}]{tannoudji}
\bibinfo{author}{\bibfnamefont{C.}~\bibnamefont{Cohen-Tannoudji}},
  \bibinfo{author}{\bibfnamefont{J.}~\bibnamefont{Dupont-Roc}},
  \bibnamefont{and} \bibinfo{author}{\bibfnamefont{G.}~\bibnamefont{Grynberg}},
  \emph{\bibinfo{title}{Atom-Photon Interactions}} (\bibinfo{publisher}{Wiley,
  New York}, \bibinfo{year}{1992}).

\bibitem[{\citenamefont{Cardimona and Stroud}(1983)}]{cardimona}
\bibinfo{author}{\bibfnamefont{D.~A.} \bibnamefont{Cardimona}}
  \bibnamefont{and} \bibinfo{author}{\bibfnamefont{C.~R.} \bibnamefont{Stroud},
  \bibfnamefont{Jr.}}, \bibinfo{journal}{Phys. Rev. A}
  \textbf{\bibinfo{volume}{27}}, \bibinfo{pages}{2456} (\bibinfo{year}{1983}).

\bibitem[{\citenamefont{Javanainen}(1992)}]{Java}
\bibinfo{author}{\bibfnamefont{J.}~\bibnamefont{Javanainen}},
  \bibinfo{journal}{Europhys. Lett.} \textbf{\bibinfo{volume}{17}},
  \bibinfo{pages}{407} (\bibinfo{year}{1992}).

\bibitem[{\citenamefont{Scully and Zubairy}(1997)}]{quantumoptics}
\bibinfo{author}{\bibfnamefont{M.~O.} \bibnamefont{Scully}} \bibnamefont{and}
  \bibinfo{author}{\bibfnamefont{M.~S.} \bibnamefont{Zubairy}},
  \emph{\bibinfo{title}{Quantum optics}} (\bibinfo{publisher}{Cambridge},
  \bibinfo{year}{1997}).

\bibitem[{\citenamefont{Dutt et~al.}(2004)\citenamefont{Dutt, Cheng, Li, Xu,
  Li, Steel, Bracker, Gammon, Economou, Liu et~al.}}]{Gurudev}
\bibinfo{author}{\bibfnamefont{M.~V.~G.} \bibnamefont{Dutt}},
  \bibinfo{author}{\bibfnamefont{J.}~\bibnamefont{Cheng}},
  \bibinfo{author}{\bibfnamefont{B.}~\bibnamefont{Li}},
  \bibinfo{author}{\bibfnamefont{X.}~\bibnamefont{Xu}},
  \bibinfo{author}{\bibfnamefont{X.}~\bibnamefont{Li}},
  \bibinfo{author}{\bibfnamefont{D.~G.~.} \bibnamefont{Steel}},
  \bibinfo{author}{\bibfnamefont{A.~S.} \bibnamefont{Bracker}},
  \bibinfo{author}{\bibfnamefont{D.}~\bibnamefont{Gammon}},
  \bibinfo{author}{\bibfnamefont{S.}~\bibnamefont{Economou}},
  \bibinfo{author}{\bibfnamefont{R.}~\bibnamefont{Liu}}, \bibnamefont{et~al.},
  \bibinfo{journal}{submitted for publication}  (\bibinfo{year}{2004}).

\bibitem[{\citenamefont{Weisskopf and Wigner}(1930)}]{weiss_wigner}
\bibinfo{author}{\bibfnamefont{V.}~\bibnamefont{Weisskopf}} \bibnamefont{and}
  \bibinfo{author}{\bibfnamefont{E.}~\bibnamefont{Wigner}},
  \bibinfo{journal}{Zeits. fur Phys.} \textbf{\bibinfo{volume}{63}}
  (\bibinfo{year}{1930}).

\bibitem[{\citenamefont{Bennett et~al.}(1996)\citenamefont{Bennett, DiVincenzo,
  Smolin, and Wootters}}]{bdsw}
\bibinfo{author}{\bibfnamefont{C.~H.} \bibnamefont{Bennett}},
  \bibinfo{author}{\bibfnamefont{D.~P.} \bibnamefont{DiVincenzo}},
  \bibinfo{author}{\bibfnamefont{J.~A.} \bibnamefont{Smolin}},
  \bibnamefont{and} \bibinfo{author}{\bibfnamefont{W.~K.}
  \bibnamefont{Wootters}}, \bibinfo{journal}{Phys. Rev. A}
  \textbf{\bibinfo{volume}{54}}, \bibinfo{pages}{3824} (\bibinfo{year}{1996}).

\bibitem[{\citenamefont{Wootters}(1998)}]{wootters}
\bibinfo{author}{\bibfnamefont{W.~K.} \bibnamefont{Wootters}},
  \bibinfo{journal}{Phys. Rev. Lett.} \textbf{\bibinfo{volume}{80}},
  \bibinfo{pages}{2245} (\bibinfo{year}{1998}).

\bibitem[{\citenamefont{Plenio et~al.}(1999)\citenamefont{Plenio, Huelga,
  Beige, and Knight}}]{plenio_cavloss}
\bibinfo{author}{\bibfnamefont{M.~B.} \bibnamefont{Plenio}},
  \bibinfo{author}{\bibfnamefont{S.~F.} \bibnamefont{Huelga}},
  \bibinfo{author}{\bibfnamefont{A.}~\bibnamefont{Beige}}, \bibnamefont{and}
  \bibinfo{author}{\bibfnamefont{P.~L.} \bibnamefont{Knight}},
  \bibinfo{journal}{Phys.\ Rev. A} \textbf{\bibinfo{volume}{59}},
  \bibinfo{pages}{2468} (\bibinfo{year}{1999}).

\bibitem[{\citenamefont{Feng et~al.}(2003)\citenamefont{Feng, Zhang, Li, Gong,
  and Xu}}]{entangle_dist_at}
\bibinfo{author}{\bibfnamefont{X.-L.} \bibnamefont{Feng}},
  \bibinfo{author}{\bibfnamefont{Z.-M.} \bibnamefont{Zhang}},
  \bibinfo{author}{\bibfnamefont{X.-D.} \bibnamefont{Li}},
  \bibinfo{author}{\bibfnamefont{S.-Q.} \bibnamefont{Gong}}, \bibnamefont{and}
  \bibinfo{author}{\bibfnamefont{Z.-Z.} \bibnamefont{Xu}},
  \bibinfo{journal}{Phys. Rev. Lett.} \textbf{\bibinfo{volume}{90}}
  (\bibinfo{year}{2003}).

\bibitem[{\citenamefont{Kim et~al.}(1999)\citenamefont{Kim, Yu, Kulik, Shih,
  and Scully}}]{quantum_eraser}
\bibinfo{author}{\bibfnamefont{Y.-H.} \bibnamefont{Kim}},
  \bibinfo{author}{\bibfnamefont{R.}~\bibnamefont{Yu}},
  \bibinfo{author}{\bibfnamefont{S.~P.} \bibnamefont{Kulik}},
  \bibinfo{author}{\bibfnamefont{Y.}~\bibnamefont{Shih}}, \bibnamefont{and}
  \bibinfo{author}{\bibfnamefont{M.~O.} \bibnamefont{Scully}},
  \bibinfo{journal}{Phys.\ Rev. Lett.} \textbf{\bibinfo{volume}{84}},
  \bibinfo{pages}{1} (\bibinfo{year}{1999}).

\bibitem[{\citenamefont{Chan et~al.}(2002)\citenamefont{Chan, Law, and
  Eberly}}]{eberly}
\bibinfo{author}{\bibfnamefont{K.~W.} \bibnamefont{Chan}},
  \bibinfo{author}{\bibfnamefont{C.~K.} \bibnamefont{Law}}, \bibnamefont{and}
  \bibinfo{author}{\bibfnamefont{J.~H.} \bibnamefont{Eberly}},
  \bibinfo{journal}{Phys.\ Rev. Lett.} \textbf{\bibinfo{volume}{88}},
  \bibinfo{pages}{100402} (\bibinfo{year}{2002}).

\bibitem[{\citenamefont{Dzhioev et~al.}(2002)\citenamefont{Dzhioev, Korenev,
  Zakharchenya, Gammon, Bracker, Tischler, and Katzer}}]{gammon}
\bibinfo{author}{\bibfnamefont{R.}~\bibnamefont{Dzhioev}},
  \bibinfo{author}{\bibfnamefont{V.}~\bibnamefont{Korenev}},
  \bibinfo{author}{\bibfnamefont{B.}~\bibnamefont{Zakharchenya}},
  \bibinfo{author}{\bibfnamefont{D.}~\bibnamefont{Gammon}},
  \bibinfo{author}{\bibfnamefont{A.}~\bibnamefont{Bracker}},
  \bibinfo{author}{\bibfnamefont{J.}~\bibnamefont{Tischler}}, \bibnamefont{and}
  \bibinfo{author}{\bibfnamefont{D.}~\bibnamefont{Katzer}},
  \bibinfo{journal}{Phys. Rev. B} \textbf{\bibinfo{volume}{66}},
  \bibinfo{pages}{153409} (\bibinfo{year}{2002}).

\bibitem[{\citenamefont{Tischler et~al.}(2002)\citenamefont{Tischler, Bracker,
  Gammon, and Park}}]{Tischler_fineStruc}
\bibinfo{author}{\bibfnamefont{J.~G.} \bibnamefont{Tischler}},
  \bibinfo{author}{\bibfnamefont{A.~S.} \bibnamefont{Bracker}},
  \bibinfo{author}{\bibfnamefont{D.}~\bibnamefont{Gammon}}, \bibnamefont{and}
  \bibinfo{author}{\bibfnamefont{D.}~\bibnamefont{Park}},
  \bibinfo{journal}{Phys. Rev. B} \textbf{\bibinfo{volume}{66}},
  \bibinfo{pages}{081310} (\bibinfo{year}{2002}).

\bibitem[{\citenamefont{Yablonovitch}(2001)}]{lighthole_yabl}
\bibinfo{author}{\bibfnamefont{E.}~\bibnamefont{Yablonovitch}},
  \bibinfo{journal}{Phys. Rev. B} \textbf{\bibinfo{volume}{64}},
  \bibinfo{pages}{125303} (\bibinfo{year}{2001}).

\bibitem[{\citenamefont{Bonadeo et~al.}(1998)\citenamefont{Bonadeo, Erland,
  Gammon, Park, Katzer, and Steel}}]{Steel_science}
\bibinfo{author}{\bibfnamefont{N.~H.} \bibnamefont{Bonadeo}},
  \bibinfo{author}{\bibfnamefont{J.}~\bibnamefont{Erland}},
  \bibinfo{author}{\bibfnamefont{D.}~\bibnamefont{Gammon}},
  \bibinfo{author}{\bibfnamefont{D.}~\bibnamefont{Park}},
  \bibinfo{author}{\bibfnamefont{D.~S.} \bibnamefont{Katzer}},
  \bibnamefont{and} \bibinfo{author}{\bibfnamefont{D.~G.} \bibnamefont{Steel}},
  \bibinfo{journal}{Science} \textbf{\bibinfo{volume}{282}},
  \bibinfo{pages}{1473} (\bibinfo{year}{1998}).

\bibitem[{\citenamefont{Yao et~al.}(2004)\citenamefont{Yao, Liu, and
  Sham}}]{qinterface}
\bibinfo{author}{\bibfnamefont{W.}~\bibnamefont{Yao}},
  \bibinfo{author}{\bibfnamefont{R.~B.} \bibnamefont{Liu}}, \bibnamefont{and}
  \bibinfo{author}{\bibfnamefont{L.~J.} \bibnamefont{Sham}},
  \bibinfo{journal}{quant-ph/0407060}  (\bibinfo{year}{2004}).

\bibitem[{\citenamefont{Bloch}(1946)}]{bloch}
\bibinfo{author}{\bibfnamefont{F.}~\bibnamefont{Bloch}},
  \bibinfo{journal}{Phys.\ Rev.} \textbf{\bibinfo{volume}{70}},
  \bibinfo{pages}{460} (\bibinfo{year}{1946}).

\bibitem[{\citenamefont{Feynman et~al.}(1957)\citenamefont{Feynman, Vernon, and
  Hellwart}}]{Feynman_spin}
\bibinfo{author}{\bibfnamefont{R.~P.} \bibnamefont{Feynman}},
  \bibinfo{author}{\bibfnamefont{F.~L.} \bibnamefont{Vernon},
  \bibfnamefont{Jr.}}, \bibnamefont{and} \bibinfo{author}{\bibfnamefont{R.~W.}
  \bibnamefont{Hellwart}}, \bibinfo{journal}{J. Appl. Phys.}
  \textbf{\bibinfo{volume}{28}}, \bibinfo{pages}{49} (\bibinfo{year}{1957}).

\bibitem[{\citenamefont{Leonhardt et~al.}(1987)\citenamefont{Leonhardt,
  Holzapfel, Zinth, and Kaiser}}]{Raman_Coherence}
\bibinfo{author}{\bibfnamefont{R.}~\bibnamefont{Leonhardt}},
  \bibinfo{author}{\bibfnamefont{W.}~\bibnamefont{Holzapfel}},
  \bibinfo{author}{\bibfnamefont{W.}~\bibnamefont{Zinth}}, \bibnamefont{and}
  \bibinfo{author}{\bibfnamefont{W.}~\bibnamefont{Kaiser}},
  \bibinfo{journal}{Chem. Phys. Letters} \textbf{\bibinfo{volume}{133}},
  \bibinfo{pages}{373} (\bibinfo{year}{1987}).

\bibitem[{\citenamefont{Bloembergen}(1996)}]{bloem}
\bibinfo{author}{\bibfnamefont{N.}~\bibnamefont{Bloembergen}},
  \emph{\bibinfo{title}{Nonlinear Optics}} (\bibinfo{publisher}{World
  Scientific, Singapore}, \bibinfo{year}{1996}).

\end{thebibliography}
\end{document}